# Capacity Constrained Influence Maximization in Social Networks

[Technical Report]

Shiqi Zhang[*]
National University of Singapore
Singapore
Southern University of Science and Technology
China
s-zhang@comp.nus.edu.sg

Yiqian Huang
Southern University of Science and Technology
China
huangyq2020@mail.sustech.edu.cn

Jiachen Sun
Tencent
China
jiachensun@tencent.com

Wenqing Lin[†]
Tencent
China
edwlin@tencent.com

Xiaokui Xiao
National University of Singapore
Singapore
xkxiao@nus.edu.sg

Bo Tang[†]
Southern University of Science and Technology
China
tangb3@sustech.edu.cn

## ABSTRACT

Influence maximization (IM) aims to identify a small number of influential individuals to maximize the information spread and finds applications in various fields. It was first introduced in the context of viral marketing, where a company pays a few influencers to promote the product. However, apart from the cost factor, the *capacity* of individuals to consume content poses challenges for implementing IM in real-world scenarios. For example, players on online gaming platforms can only interact with a limited number of friends. In addition, we observe that in these scenarios, (i) the initial adopters of promotion are likely to be the friends of influencers rather than the influencers themselves, and (ii) existing IM solutions produce sub-par results with high computational demands. Motivated by these observations, we propose a new IM variant called *capacity constrained influence maximization* (CIM), which aims to select a limited number of influential friends for each initial adopter such that the promotion can reach more users. To solve CIM effectively, we design two greedy algorithms, MG-Greedy and RR-Greedy, ensuring the 1/2-approximation ratio. To improve the efficiency, we devise the scalable implementation named RR-OPIM+ with $(1/2 - \epsilon)$-approximation and near-linear running time. We extensively evaluate the performance of 9 approaches on 6 real-world networks, and our solutions outperform all competitors in terms of result quality and running time. Additionally, we deploy RR-OPIM+ to online game scenarios, which improves the baseline considerably.

[*]This work was done while Shiqi Zhang was an intern at Tencent.
[†]Corresponding authors: Wenqing Lin and Bo Tang.

## 1 INTRODUCTION

Given a social network $G$, a diffusion model $M$ describing how a user is influenced via social connections, and a constant $k$, the influence maximization (IM) problem asks for $k$ users in $G$ that can (directly and indirectly) influence as many users as possible under $M$. It finds applications in viral marketing [11], network monitoring [25], rumor control [4], and so on. Amid them, viral marketing is the scenario where IM originates from. In this scenario, a company seeks to promote a product by incentivizing a few influential individuals in hopes of creating a cascade of adoptions via the word-of-mouth effect. Apart from the cost issue, the *individual's capacity* for spending efforts on consuming the promoting content becomes crucial when the promotion and incentive go virtual. In particular, multiple studies [34, 51] show that users have limited time to spend on social media. This leads to limited adoption of promoting content, despite its widespread distribution.

However, existing IM and its variants neglect the capacity of individuals, hindering their practicalities in the corresponding marketing scenario. Before explaining, we first introduce incentive propagation events of online games, whose objective is stimulating interactions between friends. Specifically, the service provider of online gaming platforms distributes virtual incentives to a set of pre-selected users, called *active participants* (APs), who are more likely to participate in this promotion, and encourages them to play with friends from a recommendation list. Once the recommended friends play with APs, they also attain the incentive and can share it with their friends during daily co-playing. The crux of in-game incentive propagation events is recommending existing friends to APs. This is because each user has numerous friends but only has a limited capacity to play with them. In addition, we observe that the APs are more likely to be *the friends of influencers rather than the influencers themselves*, highlighting the importance of choosing influential friends for APs. In contrast, IM and most of its variants assume that each selected influencer unconditionally adopts the promotion from the merchant without relying on friends, which contradicts this insight. Moreover, it is rather challenging to utilize IM and corresponding solutions [22, 46, 49], since independently selecting friends for each AP by the IM solver can incur immense computational overhead, and the result quality remains unclear.

To this end, we propose a new IM variant called the *capacity constrained influence maximization* (CIM) problem. Given a social network $G$, a diffusion model M, a set $A$ of APs and a constant $k$, CIM aims to find $k$ influential friends (seeds) for each user in $A$, such that the number of influenced users starting from all distinct seeds is maximized under M. To solve CIM, we design a vanilla solution MG-Greedy, which employs a greedy strategy to select the best feasible seed from all neighbors of $A$ and provides a 1/2-approximation guarantee. In addition, we propose the solution RR-Greedy to select seeds for each user of $A$ in a round-robin manner, which improves the time complexity of MG-Greedy by a factor of $|A|$ and ensures at least 1/2-approximation. To improve the efficiency, we further propose the scalable implementation RR-OPIM+, which shares the same framework with the state-of-the-art IM solution OPIM-C [46] but is redesigned carefully to ensure the correctness for CIM. Most notably, RR-OPIM+ achieves $(1/2 - \epsilon)$-approximation in a near-linear running time w.r.t. the network scale.

In experiments, we first provide an empirical configuration for CIM based on incentive propagation events. Subsequently, we extensively evaluate the performance of 9 approaches on 6 real-world networks with up to 3 billion relationships. Notably, our proposals outperform all competitors by up to 39% in terms of result quality. Besides, RR-OPIM+ speeds up greedy algorithms by at least 4 orders of magnitude. In addition, we deploy our solution RR-OPIM+ to the online gaming scenario, which improves the baseline by up to 5.39% in the corresponding evaluation metric.

To summarize, we make the following contributions in this work:

- We conduct an empirical study to verify the difference between in-game incentive propagation and IM. Motivated by these observations, we propose a new IM variant called CIM. (Section 3)
- For effectiveness, we propose two CIM solutions MG-Greedy and RR-Greedy with approximation guarantees. (Section 4)
- For efficiency, we provide a scalable implementation with rigorous theoretical analysis for these greedy algorithms. (Section 5)
- We discover the detailed settings for CIM and conduct experiments to show the superiority of our proposals. (Section 7)
- We deploy the proposal to the in-game incentive propagation event, which achieves considerable improvement. (Section 8)

## 2 PRELIMINARIES

We abstract a social network as a graph $G = (V, E)$, where $V$ is a set of $n$ nodes (representing users) and $E$ is a set of $m$ edges (representing relationships). We assume that $G$ is a directed graph and each edge $e_{u,v} \in E$ indicates that $v$ is a follower of and can be influenced by $u$. We call $u$ (resp. $v$) the in-neighbor (resp. out-neighbor) of $v$ (resp. $u$). Furthermore, we use $N_u$ to denote the set of out-neighbors of $u$. For an undirected graph, we replace each undirected edge $e_{u,v}$ with two directed ones in opposing directions, i.e., $e_{u,v}$ and $e_{v,u}$. In the sequel, we elaborate on the background of influence maximization (IM), followed by in-game incentive propagation scenarios.

### 2.1 Diffusion Models

This work focuses on two well-accepted diffusion models, named Independent Cascade (IC) [14] and Linear Threshold (LT) [15]. Given a graph $G$ and a set $S$ of chosen users (called seeds), both models assume that each user $u \in V$ has two possible states: inactive or active, and describe the diffusion of an item from $S$ in a stochastic manner. Specifically, the states of seeds are set to be active at the initial step $t = 0$, and at step $t > 0$, the newly activated users try to influence their inactive out-neighbors as follows.

- IC introduces the influence probability $p_{u,v}$ for each edge $e_{u,v}$, representing the likelihood that $v$ is successfully activated by $u$. At each step $t > 0$, each user $u$ who is activated as step $t - 1$ has *one chance* to influence each inactive out-neighbor $v$ with $p_{u,v}$.
- LT assumes that each edge $e_{u,v}$ is associated with a weight $w_{u,v}$ satisfying $\sum_{u \in N_v^{in}} w_{u,v} \leq 1$, where $N_v^{in}$ is the set of in-neighbors of $v$. At $t = 0$, the threshold $\theta_v \in [0, 1]$ is uniformly sampled for each user $v$. For $t > 0$, an inactive user $v$ is activated if $\phi(v, t)$ exceeds the threshold $\theta_v$, where $\phi(v, t)$ is the summation of $w_{u,v}$ w.r.t. $v$'s in-neighbor $u$ that was activated before step $t$.

A user remains active in all subsequent steps once it is activated. The influence process continues until no more users can be activated.

In this paper, M is denoted as the diffusion model IC or LT, and influence spread $\sigma_{G,M}(S)$ is defined as the expected number of nodes on $G$ activated by the set $S$ under the model M. We denote $\sigma_{G,M}(v|S) = \sigma_{G,M}(S \cup \{v\}) - \sigma_{G,M}(S)$ as the *marginal gain* of adding $v$ to a set $S$. As an instance of the non-decreasing submodular function, $\sigma_{G,M}(\cdot)$ satisfies the following properties.

PROPOSITION 1 ([22]). *The spread $\sigma_{G,M}(\cdot)$ is a non-decreasing submodular set function satisfying: (i) $0 \leq \sigma_{G,M}(S) \leq \sigma_{G,M}(T), \forall S \subseteq T \subseteq V$; (ii) $\sigma_{G,M}(v|S) \geq \sigma_{G,M}(v|T), \forall S \subseteq T \subseteq V$ and $v \in V \setminus T$.*

### 2.2 Influence Estimation and Maximization

Given a graph $G$, a model M, and a constant $k$, the influence maximization (IM) problem [11, 22, 39] asks for a seed set $S$ with cardinality $|S| = k$ such that the influence spread $\sigma_{G,M}(S)$ is maximized. Kempe et al. [22] formulate this problem and provide a $(1 - 1/e)$−approximate IM solution if the exact spread $\sigma_{G,M}(S)$ is known for any $S$. Due to the #P-hardness [6] of evaluating $\sigma_{G,M}(S)$, most of previous solutions [3, 17, 18, 22, 25, 46, 47, 49, 50, 58] employ the Monte-Carlo (MC) simulation [22] or Reverse-Reachable (RR) set sampling [3] to estimate the spread, yielding a $(1 - 1/e - \epsilon)$−approximate result. In what follows, we briefly review these two estimation techniques and representative solutions based thereon.

**MC simulation.** Given a graph $G$, a model M, and a node set $S$, the MC simulation starts from $S$ and estimates $\sigma_{G,M}(S)$ following the discrete step of M in Section 2.1. To reduce the estimation variance, the simulation is conducted $r$ times, and the average number of activated nodes in $r$ trials is recorded as the estimation of $\sigma_{G,M}(S)$. Kempe et al. [22] propose a greedy solution based on MC simulations, which iteratively includes a node $v$ into $S$ as the $i$-th seed if the marginal gain $\sigma_{G,M}(v|S)$ is the largest. To avoid evaluating $\sigma_{G,M}(v|S)$ for $O(n)$ nodes in each iteration, Leskovec et al. [25] exploit the submodularity and propose a practical implementation called CELF, which skips $v$ while selecting the $j$-th seed if $v$'s marginal gain is sufficiently small in the iteration prior to $j$.

**RR set sampling.** Borgs et al. [3] propose to estimate the spread by sampling random RR sets, defined as follows.

*Definition 2.1 (RR Set).* Given a graph $G$ and a model M, a random RR set $R_{G,M}$ is a set of nodes, generated by (i) first selecting a node $v$ at random, (ii) then sampling a subgraph $g$ from $G$ in terms of M, (iii) finally preserving the nodes that can reach $v$ in $g$. $\mathcal{R}_{G,M}$ is denoted as a set of random RR sets.

In Definition 2.1, the subgraph $g$ is sampled based on the influence process of M. For IC, $g$ is induced by removing each edge $e_{u,v}$ in $G$ with the probability $1 - p_{u,v}$. Borgs et al. [3] prove that $\sigma_{G,M}(S) = n \cdot \Pr[S \cap R_{G,M} \neq \emptyset]$. In other words, $n \cdot \frac{\Lambda_{\mathcal{R},G,M}(S)}{\theta}$ is an unbiased estimator of $\sigma_{G,M}(S)$, where $\mathcal{R}_{G,M}$ is a set of $\theta$ random RR sets and $\Lambda_{\mathcal{R},G,M}(S)$ is the coverage of $S$ in $\mathcal{R}_{G,M}$, i.e., the number of RR set $R_{G,M} \in \mathcal{R}_{G,M}$ satisfying $S \cap R_{G,M} \neq \emptyset$. By this connection, Borgs et al. [3] sample a sufficient number of random RR sets as $\mathcal{R}_{G,M}$, and employ the greedy framework in [22] to iteratively select the next node $v$ with the largest *marginal coverage*

$$\Lambda_{\mathcal{R},G,M}(v|S) = \Lambda_{\mathcal{R},G,M}(S \cup \{v\}) - \Lambda_{\mathcal{R},G,M}(S).$$

The related solutions [17, 18, 46, 49, 50] follow the greedy strategy in [3] and improve its efficiency by reducing $\theta$ while ensuring the same approximation ratio. At this front, OPIM-C [46] is state of the art and is applied to subsequent IM solutions and variants [2, 17, 18].

In addition, a line of solutions [9, 35] estimates spread by sampling graph instances, and a line of solutions either leverages centrality scores [7, 21] or simplifies the diffusion model [6] to generate seed heuristically. We refer interested readers to [28] for details.

### 2.3 In-Game Incentive Propagation

**Event procedure.** The online gaming platform designs the incentive propagation event to boost user engagement, whose procedures are as follows. Given a social network $G = (V, E)$, the service provider first selects a set of users, named *active participants* (APs), who are more likely to engage in this event. We denote this set of APs as $A$ with $|A| = d \ll n$. We say a user $v$ is a *passive participant* (PP) if $v \in V \backslash A$. After $A$ is chosen, the service provider then distributes this event to $A$, including the mission details, the incentive $T$ (e.g., extra credits), and a list of PP friends (named *passive seeds*). $T$ is initially possessed by $A$ and is shared by the following two steps, where we say a user is *activated* if attaining $T$.

- *Seed activation.* Each AP $u$ is the initial active user and can invite the recommended seed $v$, who becomes active automatically after playing with one of AP inviters for the first time.
- *Daily contamination.* Starting from activated seeds, $T$ can be recursively shared from an active PP to an inactive PP friend if they play together for the first time during the event.

Notice that the provider can recommend at most $k$ passive seeds for each AP, since a user has a large number of friends (from gaming and messaging platforms) but only has a limited capacity to play with them. The passive seed set for each AP is defined as follows.

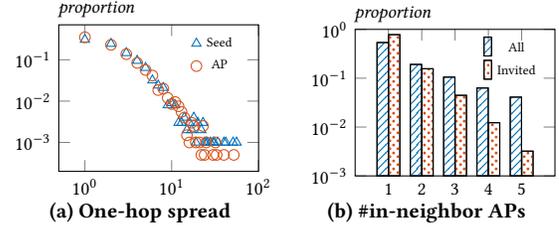

**Figure 1: The distribution of one-hop spread of APs and their seeds and the number of in-neighbor APs of seeds on *TXG*.**

*Definition 2.2 (Passive Seed Set).* Given a graph $G$, an AP set $A$, an AP $u$, and a constant $k$, let $C_u = N_u \backslash A$ be the set of $u$'s passive out-neighbors. The passive seed set of $u$ is $S_u \subseteq C_u$ with $|S_u| \leq k$.

**Existing and possible selection approaches.** The linchpin of the incentive propagation event is selecting $k$ seeds for each AP. Existing strategies can be generalized to a local framework, which independently ranks the passive friends for each AP $u \in A$ in descending order based on a heuristic score, e.g., degree, PageRank, the number of historical interactions, and so on [20, 29, 30, 32, 36], and select top $k$ friends with the highest scores as seeds. To improve the number of engaged users, the idea of IM could be utilized by invoking an IM solver as mentioned in Section 2.2 and selecting $k$ seeds from $C_u$ for each AP $u$. However, this local framework yields numerous *overlapping seeds*, each of which is assigned to more than one APs, rendering the compromised spread. In the meantime, although the IM solver can output an approximate seed set w.r.t. $C_u$ of each AP $u$, the quality of the overall seed set is unclear.

## 3 THE CIM PROBLEM

In this part, we first conduct an empirical study to (i) clarify the difference between IM and the in-game incentive propagation scenario (Observation 1), and (ii) verify the drawback of the local framework (Observation 2). Motivated by these observations, we then formulate capacity constrained influence maximization (CIM).

### 3.1 Motivating Insights

We collect user logs from an incentive propagation event of a Tencent role-playing game, which follows the procedure in Section 2.3, and call this dataset *TXG*. In particular, the service provider selects the monthly active user as the AP, who is randomly assigned one of the following strategies: PageRank, degree, or the number of historical interactions, to obtain at most $k = 40$ recommended friends. *TXG* consists of (i) a social network $G$ with 243.4 thousand users and 5.9 million undirected friendships, (ii) 7.6 thousand APs and 3.9 thousand seeds involved in seed activation, and (iii) 18.5 thousand PPs activated in daily contamination.

In the first set of observations, we explore whether influencers tend to participate in the event through the service provider or via invitations from friends. To reduce the bias during the exploration of user influence, we excluded APs and their seeds recommended by the influence-based strategy degree or PageRank. For each AP and seed, we define the one-hop spread as the number of seeds that played with each AP in seed activation and the number of PPs that are directly activated by each seed in daily contamination, respectively. Figure 1(a) reports the distribution of the one-hop spread of the rest APs and their seeds, providing the insight below.

OBSERVATION 1 (LESS-INFLUENTIAL APs). *The APs of in-game incentive propagation events are less influential than their seeds.*

Specifically, the average one-hop spread of seeds is 7% larger than that of APs, and the fraction of passive friends with a one-hop spread larger than 50 is 4× more than APs.

To demonstrate the existence of overlapping seeds in existing strategies, Figure 1(b) reports the distribution of the number of in-neighbor APs of each seed and the distribution of the number of APs who have invited this seed. We call a seed an overlapping seed if it has more than one in-neighbor AP. As shown in Figure 1(b), since seeds are independently recommended to each AP, 46.3% fraction of seeds are overlapping seeds, which incurs 22.1% fraction of passive seeds invited by more than one APs. However, the engagement likelihood of a seed is weakly related to the frequency of being invited. Specifically, for seeds invited once, twice, and third times, the fraction of engaged passive seeds is 41.5%, 45.4%, and 42.6%, respectively. This leads to the second observation.

OBSERVATION 2 (OVERLAPPING SEEDS). *Due to the overlapped neighborhood and the independent recommendation strategy, the passive seed can be invited by multiple APs, however, the repeated invitation has a slight impact on the engagement willingness of seeds.*

To ensure the robustness of our findings, we expand our exploration to two additional in-game incentive propagation events, which confirm our previous observations. These observations can also find support from Epinions [33], Twitter[1], and Facebook[8].

### 3.2 Problem Formulation

Prior to CIM, we first define the spread of passive participants.

*Definition 3.1 (Spread of Passive Participants).* Given a graph $G = (V, E)$, a model M, and an AP set $A$, let $P = (V_p, E_p)$ be the graph of passive participants, where $V_p = V \setminus A$ and $E_p = \{e_{u,v} \in E : u \notin A, v \notin A\}$. For any node set $S = \bigcup_{u \in A} S_u \subset V_p$, the spread of passive participants $\sigma_{P,M}(S)$ is defined as the expected number of nodes on $P$ activated by $S$ under M.

Distinct from the spread $\sigma_{G,M}(S)$ in IM, the spread of PPs is measured on the induced subgraph $P$. This is due to that the incentive propagation event treats APs as active users prior to seeds, and the influence paths from seeds to other PPs via any APs are invalid by rules. Notice that the individual capacity for PP's interactions is considered into M by leveraging the historical interaction logs, which will be explained in Section 7.2. To address the overlapping issue in Observation 2, we evaluate the *union* of seeds of all APs, which intuitively recommends more distinct seeds and can fully leverage the recommendation list. For ease of presentation, we omit the subscript of $\sigma_{P,M}(\cdot)$ unless otherwise specified. We now introduce CIM in Problem 1, which is NP-hard as shown in Corollary 1. For ease of exposition, we defer all proofs to Appendix A.

PROBLEM 1 (CAPACITY CONSTRAINED INFLUENCE MAXIMIZATION (CIM)). *Given a graph $G$, a model M, an AP set $A$, and a constant $k$, CIM aims to find the optimal passive seeds set $S_u^*$ for each $u \in A$ such that the spread $\sigma\left(\bigcup_{u \in A} S_u^*\right) = \sigma(S^*)$ is maximized.*

COROLLARY 1. *The CIM problem is NP-hard.*

Besides the incentive propagation scenario of online games, CIM can be applied to more complex viral marketing scenarios, such as advertising to different cohorts of users [16]. In addition, it is worth noting that CIM only identifies influential friends for APs. The interaction inclination between the AP and influential friend is orthogonal to this work, which can be directly applied to our proposal and will be explained in Section 8.

## 4 GREEDY ALGORITHMS

In this section, we first introduce the main idea of our proposal, and then propose two greedy solutions, named Maximal Gain Greedy (MG-Greedy) and Round-Robin Greedy (RR-Greedy), including the correctness and complexity analysis. We assume that each greedy solution has an *oracle* to access the *exact* spread value $\sigma(\cdot)$ and we focus on the number of times accessing the exact spread in the complexity analysis. Regarding the estimation and its overhead of $\sigma(\cdot)$, we defer them to Section 5.

### 4.1 Main Idea

Since the spread of PPs is a non-decreasing submodular function, we solve CIM by converting it to one instance of submodular maximization under *partition matroid* [12], whose concept is as follows.

*Definition 4.1 (Partition Matroid [12]).* Given a set $U$, disjoint sets $U_1, \ldots, U_d$ with $\bigcup_{i \in [d]} U_i = U$ and constants $k_1, \ldots, k_d$, the partition matroid of $U$ is $(U, \{X \subseteq U : |X \cap U_i| \leq k_i, \forall i \in [d]\})$.

Different from partitions $U_1, \ldots, U_d$ in Definition 4.1, the passive friend set $C_u$ for each $u \in A$ is not strictly disjoint in CIM. To this end, we use the calligraphic notation $C_u = \{e_{u,v} : v \in C_u\}$ (resp. $S_u = \{e_{u,v} : v \in S_u\}$) to represent the set of distinct edges from AP $u$ to its passive friends (resp. passive seeds). Analogously, we let $C = \bigcup_{u \in A} C_u$ and $S = \bigcup_{u \in A} S_u$ be the global candidate space of AP-PP edges for seed selection and the set of selected AP-seed edges, respectively. Hence, CIM turns to find $S^* \in I$ such that $\sigma(S^*)$ is maximized, where the partition matroid $(C, I)$ is $I = \{S : |S_u| \leq k, \forall u \in A\}$. In what follows, we use the calligraphic notation $S$, where $\sigma(S) = \sigma(S)$ and $\sigma(e_{u,v}|S) = \sigma(v|S)$.

### 4.2 MG-Greedy and RR-Greedy

**MG-Greedy.** We design a greedy solution MG-Greedy, extending the greedy solution of IM [22]. Intuitively, MG-Greedy selects the edge with the largest marginal gain from the global candidate space $C$ while maintaining the partition matroid. As illustrated in Algorithm 1, at each iteration (Lines 3-5), MG-Greedy selects the edge $e_{u,v}$ with the largest marginal gain $\sigma(e_{u,v}|S)$ from $C$, where ties are settled arbitrarily. If $S_u$ is feasible (i.e., $|S_u| < k$), the selected $e_{u,v}$ is then added to $S$; otherwise, all edges starting from $u$ are removed from $C$. MG-Greedy accesses the spread function $O(d \cdot k \cdot |C|)$ times, containing $O(|C|)$ times of spread function invocations in each of $O(d \cdot k)$ iterations. As proved by [12], MG-Greedy has the following guarantee when the spread oracles are given.

THEOREM 4.2. *Given a graph $G$, a model M, an AP set $A$, and a constant $k$, let $S^*$ be the optimal solution of CIM. The output $S$ of MG-Greedy satisfies $\sigma(S) \geq \frac{1}{2} \cdot \sigma(S^*)$.*

---

**Algorithm 1:** MG-Greedy $(G, A, k, \sigma)$

1　$\mathcal{S} \leftarrow \emptyset; C \leftarrow \bigcup_{u \in A} C_u; t_u \leftarrow 0, \forall u \in A$;
2　**while** $C \setminus \mathcal{S} \neq \emptyset$ **do**
3　　　$e_{u,v} \leftarrow \arg\max_{e_{u',v'} \in C \setminus \mathcal{S}} \sigma(e_{u',v'}|\mathcal{S})$;
4　　　$\mathcal{S} \leftarrow \mathcal{S} \cup \{e_{u,v}\}; t_u \leftarrow t_u + 1$;
5　　　**if** $t_u \geq k$ **then** $C \leftarrow C \setminus C_u$;
6　**return** $\mathcal{S}$;

---

**RR-Greedy.** Notice that MG-Greedy yields a 1/2-approximate output by iteratively selecting only one AP-seed pair after evaluating the marginal gain of $O(|C|)$ pairs. To improve the result quality and reduce invocation times of spread functions, we propose a greedy algorithm called RR-Greedy, ensuring an at least 1/2-approximate result. The main idea is selecting the edge from the local candidate space $C_u$ for each AP $u$. Let $L \subseteq A$ be the set where each $u \in L$ has selected less than $k$ seeds. As illustrated in Algorithm 2, at each iteration, RR-Greedy selects the edge $e_{u,v}$ with the largest marginal gain $\sigma(e_{u,v}|\mathcal{S})$ from $C_u$ for each $u \in L$ (Lines 3-4), where ties are settled arbitrarily. RR-Greedy adds the selected $e_{u,v}$ to $\mathcal{S}$, and removes $u$ from $L$ if it has enough seeds or no more candidates exist (Lines 5-6). RR-Greedy accesses the spread oracle $O(k \cdot |C|)$ times, improving MG-Greedy by $O(d)$, and its correctness is as follows.

**Theorem 4.3.** *Given a graph $G$, a model $M$, an AP set $A$, and a constant $k$, let $\mathcal{S}^*$ be the optimal solution of* CIM. *The output $\mathcal{S}$ of RR-Greedy satisfies $\sigma(\mathcal{S}) \geq \frac{1}{1+\gamma} \cdot \sigma(\mathcal{S}^*)$, where $\gamma$ is a constant in $[0, 1]$ satisfying $\sigma(e_{u,v}|\mathcal{T}) \geq (1 - \gamma) \cdot \sigma(e_{u,v}|\mathcal{S})$, for all $\mathcal{S} \subseteq \mathcal{T}$ that $|(\mathcal{T} \setminus \mathcal{S}) \cap C_u| \leq k \; \forall u \in A$ and $e_{u,v} \in C \setminus \mathcal{T}$.*

Hence, RR-Greedy is $\frac{1}{1+\gamma}$-approximate, which is superior over MG-Greedy. The constant $\gamma$ in Theorem 4.3 is also known as *curvature* [13] and is further bounded by

$$\gamma \leq \gamma_{max} = 1 - \min_{e_{u,v} \in C} \frac{\sigma(C) - \sigma(C \setminus \{e_{u,v}\})}{\sigma(\{e_{u,v}\})}.$$

## 5 SCALABLE IMPLEMENTIONS

To estimate the spread, we propose a scalable implementation RR-OPIM+ for RR-Greedy, extending the state-of-the-art OPIM-C [46]. In the sequel, we introduce the main idea of it and clarify its difference from OPIM-C, followed by its implementation and analysis.

### 5.1 Main Idea

Given a graph $G$, an AP set $A$ and the induced subgraph $P$ with $n_p$ nodes, let $\mathcal{R}_{P,M}$ be a set of random RR sets constructed from $P$. Akin to Section 2.2, the connection between a random RR set $R_{P,M} \in \mathcal{R}_{P,M}$ and the spread $\sigma_{P,M}(\cdot)$ is $\sigma_{P,M}(S) = n_p \cdot \Pr[S \cap R_{P,M} \neq \emptyset]$. Hence, the objective in CIM can be solved by finding the seed $S$ with the maximum coverage $\Lambda_{\mathcal{R}_{P,M}}(S)$ in $\mathcal{R}_{P,M}$, and the greedy solutions in Section 4 can be efficiently implemented by replacing the evaluation of $\sigma_{P,M}(v|S)$ with $\Lambda_{\mathcal{R}_{P,M}}(v|S)$. In analogy to $\sigma_{P,M}(\cdot)$, we omit the subscripts $P$ and $M$ of $\mathcal{R}_{P,M}(\cdot)$ and $\Lambda_{\mathcal{R}_{P,M}}(\cdot)$ in the following contexts. Furthermore, the coverage $\Lambda_{\mathcal{R}}(S) = \Lambda_{\mathcal{R}}(\mathcal{S})$ and marginal coverage $\Lambda_{\mathcal{R}}(v|S) = \Lambda_{\mathcal{R}}(e_{u,v}|\mathcal{S})$ in the edge notation.

---

**Algorithm 2:** RR-Greedy $(G, A, k, \sigma)$

1　$\mathcal{S} \leftarrow \emptyset; L = A; t_u \leftarrow 0, \forall u \in A$;
2　**while** $L \neq \emptyset$ **do**
3　　**foreach** $u \in L$ **do**
4　　　　$e_{u,v} \leftarrow \arg\max_{e_{u',v'} \in C_u \setminus \mathcal{S}} \sigma(e_{u',v'}|\mathcal{S})$;
5　　　　$\mathcal{S} \leftarrow \mathcal{S} \cup \{e_{u,v}\}; t_u \leftarrow t_u + 1$;
6　　　　**if** $t_u \geq k$ **or** $C_u \setminus \mathcal{S} = \emptyset$ **then** $L = L \setminus \{u\}$ ;
7　**return** $\mathcal{S}$;

---

The pseudocode of RR-OPIM+ is illustrated in Algorithm 3. Initially, RR-OPIM+ defines the constants $\theta_{max}$ and $\theta$ (Lines 1-2), representing the worst-case and the initial number of random RR sets, respectively, and then constructs two sets $\mathcal{R}_1$ and $\mathcal{R}_2$, both containing $\theta$ random RR sets (Line 3). After that, it runs in an iterative manner to verify if the selected seed pairs have satisfied the approximation guarantee by using less than $\theta_{max}$ random RR sets. At each iteration, it first invokes RR-Greedy by using $\Lambda_{\mathcal{R}_1}(\cdot)$ as the evaluation function and selects the set $\mathcal{S}$ of AP-seed edges (Line 5). To verify if the selected $\mathcal{S}$ provides the desired approximation guarantee, it then computes the upper bound $\sigma^u(\mathcal{S}^*)$ of $\sigma(\mathcal{S}^*)$ (resp. the lower bound $\sigma^l(\mathcal{S})$ of $\sigma(\mathcal{S})$) by using $\mathcal{R}_1$ (resp. $\mathcal{R}_2$) (Lines 6-7). RR-OPIM+ is early terminated with current $\mathcal{S}$ if

$$\frac{\sigma(\mathcal{S})}{\sigma(\mathcal{S}^*)} \geq \frac{\sigma^l(\mathcal{S})}{\sigma^u(\mathcal{S}^*)} \geq \frac{1}{2} - \epsilon.$$

Or, it doubles the sizes of $\mathcal{R}_1$ and $\mathcal{R}_2$ and continues (Lines 8-9).

Although RR-OPIM+ and OPIM-C [46] share the framework as shown in Algorithm 3, there remain three challenges while considering CIM and RR-Greedy: (i) a new $\theta_{max}$ requires devising to ensure the approximation guarantee in the worst case; (ii) $\sigma^u(\mathcal{S}^*)$ and $\sigma^l(\mathcal{S})$ are to recompute to secure the approximation for any of $i_{max}$ early terminations; (iii) the result returned from any of above-said criteria is correct with a high probability. To address these issues, we implement the following three major modifications.

### 5.2 Detailed Modifications

**Computing $\theta_{max}$.** As shown in Line 1 of Algorithm 3, RR-OPIM+ first generates a random seed set $\mathcal{S}$ by assigning each candidate PP to an arbitrary AP friend while ensuring the partition matroid $\mathcal{I}$. It then records the number of distinct seeds in $\mathcal{S}$ as $\chi$, which is a lower bound of the optimal spread $\sigma(\mathcal{S}^*)$ and is required for deriving $\theta_{max}$. The following lemma provides the setting of $\theta_{max}$, ensuring the correctness of RR-OPIM+ when $i = i_{max}$.

**Lemma 5.1.** *Let $\mathcal{R}$ be a set of random RR sets, $\chi$ be defined as Line 1, and $\mathcal{S}$ be the result obtained by Line 8 of Algorithm 3. For fixed $\epsilon$ and $\delta$, if $|\mathcal{R}| \geq \theta_{max}$ and*

$$\theta_{max} = \frac{2n_p \cdot \left(\frac{1}{2}\sqrt{\ln \frac{6}{\delta}} + \sqrt{\frac{1}{2} \cdot \left(\ln\left(\prod_{u \in A} \binom{|C_u|}{k}\right) + \ln \frac{6}{\delta}\right)}\right)^2}{\epsilon^2 \cdot \chi}, \quad (1)$$

*then $\mathcal{S}$ is $(1/2 - \epsilon)$-approximate with at least $1 - \delta/3$ probability.*

**Bounding $\sigma(\mathcal{S}^*)$ and $\sigma(\mathcal{S})$.** We next derive the lower bound $\sigma^l(\mathcal{S})$ of $\sigma(\mathcal{S})$ and the upper bound $\sigma^u(\mathcal{S}^*)$ of $\sigma(\mathcal{S}^*)$ such that the approximation ratio $\frac{\sigma(\mathcal{S})}{\sigma(\mathcal{S}^*)} \geq \frac{\sigma^l(\mathcal{S})}{\sigma^u(\mathcal{S}^*)}$. The settings are as follows.

---

**Algorithm 3:** RR-OPIM+ $(G, A, k, \epsilon, \delta)$

1  Let $\mathcal{S} \in \mathcal{I}$ be a random seed set with $\chi$ distinct seeds;
2  $\theta_{max} \leftarrow$ Eq. (1); $\theta \leftarrow \frac{\epsilon^2}{n_p} \cdot \theta_{max}$; $i_{max} \leftarrow \left\lceil \log_2 \frac{\theta_{max}}{\theta} \right\rceil$;
3  Generate $\mathcal{R}_1$ and $\mathcal{R}_2$ with $|\mathcal{R}_1| = |\mathcal{R}_2| = \theta$;
4  **for** $i = 1, 2, \ldots, i_{max}$ **do**
5  $\quad$ $\mathcal{S} \leftarrow$ RR-Greedy $(G, A, k, \Lambda_{\mathcal{R}_1})$;
6  $\quad$ Compute $\sigma^u(\mathcal{S}^*)$ by Eq.(4) on $\mathcal{R}_1$ with $p_f = \frac{\delta}{3 \cdot i_{max}}$;
7  $\quad$ Compute $\sigma^l(\mathcal{S})$ by Eq.(3) on $\mathcal{R}_2$ with $p_f = \frac{\delta}{3 \cdot i_{max}}$;
8  $\quad$ **if** $\frac{\sigma^l(\mathcal{S})}{\sigma^u(\mathcal{S}^*)} \geq \frac{1}{2} - \epsilon$ **or** $i = i_{max}$ **then return** $\mathcal{S}$;
9  $\quad$ Double the sizes of $\mathcal{R}_1$ and $\mathcal{R}_2$ with new random RR sets;

---

LEMMA 5.2. *Given a graph $P$ with $n_p$ nodes, $\mathcal{R}_1$ with $|\mathcal{R}_1| = \theta_1$, and $\mathcal{R}_2$ with $|\mathcal{R}_2| = \theta_2$, for any $p_f \in (0, 1)$, by setting*

$$\sigma^u(\mathcal{S}^*) = \left( \sqrt{2 \cdot \Lambda_{\mathcal{R}_1}(\mathcal{S}) + \frac{\ln(1/p_f)}{2}} + \sqrt{\frac{\ln(1/p_f)}{2}} \right)^2 \cdot \frac{n_p}{\theta_1}, \quad (2)$$

$$\sigma^l(\mathcal{S}) = \left( \left( \sqrt{\Lambda_{\mathcal{R}_2}(\mathcal{S}) + \frac{2 \cdot \ln(1/p_f)}{9}} - \sqrt{\frac{\ln(1/p_f)}{2}} \right)^2 - \frac{\ln(1/p_f)}{18} \right) \cdot \frac{n_p}{\theta_2}, \quad (3)$$

*we have* $\Pr\left[ \sigma(\mathcal{S}) < \sigma^l(\mathcal{S}) \right] < p_f$ *and* $\Pr\left[ \sigma(\mathcal{S}^*) > \sigma^u(\mathcal{S}) \right] < p_f$.

In Lemma 5.2, the derivation of $\sigma^u(\mathcal{S}^*)$ requires an upper bound of the coverage $\Lambda_{\mathcal{R}_1}(\mathcal{S}^*)$ of the unknown optimal set $\mathcal{S}^*$. To this end, we need the following corollary in terms of Theorem 4.3, since $\Lambda_{\mathcal{R}}(\cdot)$ is also non-decreasing and submodular.

COROLLARY 2. *Let $\mathcal{S}^*$ be the optimal solution of* CIM. *RR-Greedy with $\Lambda_{\mathcal{R}}(\cdot)$ outputs an $\mathcal{S}$ with $\Lambda_{\mathcal{R}}(\mathcal{S}) \geq \frac{1}{1+\gamma} \Lambda_{\mathcal{R}}(\mathcal{S}^*) \geq \frac{1}{2} \Lambda_{\mathcal{R}}(\mathcal{S}^*)$.*

Accordingly, Lemma 5.2 employs $2 \cdot \Lambda_{\mathcal{R}_1}(\mathcal{S})$ as a vanilla upper bound of $\Lambda_{\mathcal{R}_1}(\mathcal{S}^*)$, which might be loose in practice and motivates us to design a tightened upper bound of $\Lambda_{\mathcal{R}_1}(\mathcal{S}^*)$ as follows.

LEMMA 5.3. *For any seed set $\mathcal{S}$ under partition matroid, i.e., $|\mathcal{S}_u| \leq k, \forall u \in A$ and any set $\mathcal{R}$ of random RR sets,*

$$\Lambda_{\mathcal{R}}(\mathcal{S}^*) \leq \Lambda^\phi_{\mathcal{R}}(\mathcal{S}^*) = \Lambda_{\mathcal{R}}(\mathcal{S}) + \sum_{u \in A} \sum_{e_{u,v} \in \Phi_k(\mathcal{S}, u)} \Lambda_{\mathcal{R}}(e_{u,v}|\mathcal{S}),$$

*where $\Phi_k(\mathcal{S}, u)$ denotes the set of at most $k$ AP-seed pairs in $C_u$ with the $k$ largest coverage gain on $\mathcal{R}$ w.r.t. $\mathcal{S}$.*

Accordingly, a tightened upper bound $\sigma^u(\mathcal{S}^*)$ is

$$\sigma^u(\mathcal{S}^*) = \left( \sqrt{\Lambda^u_{\mathcal{R}_1}(\mathcal{S}^*) + \frac{\ln 1/p_f}{2}} + \sqrt{\frac{\ln 1/p_f}{2}} \right)^2 \cdot \frac{n_p}{\theta_1}, \quad (4)$$

where

$$\Lambda^u_{\mathcal{R}_1}(\mathcal{S}^*) = \min \left\{ 2 \cdot \Lambda_{\mathcal{R}_1}(\mathcal{S}), \min_{0 \leq t < k} \Lambda^\phi_{\mathcal{R}_1}(\mathcal{S}^t) \right\}$$

and $\mathcal{S}^t$ is the seed set with $|\mathcal{S}^t \cap C_u| = \min(|C_u|, t) \, \forall u \in A$.

**Putting it together.** As per Algorithm 3 and Lemma 5.2, by setting $p_f = \delta/(3 \cdot i_{max})$, we derive an unqualified $\sigma^l(\mathcal{S}) > \sigma(\mathcal{S})$ (or $\sigma^u(\mathcal{S}) < \sigma(\mathcal{S}^*)$) with probability at most $\delta/(3 \cdot i_{max})$ in a given iteration $i$ of $i_{max}$ iterations. Moreover, as illustrated in Lemma 5.1, the failure probability for the result in the worst case is at most $\delta/3$. By the union bound, the correctness of RR-OPIM+ is as follows.

THEOREM 5.4. *Given a graph $G$, a set of APs $A$, a model $M$, and a constant $k$, let $\mathcal{S}^* = \bigcup_{u \in A} \mathcal{S}^*_u$ be the optimal solution of* CIM. *For every $\epsilon > 0$ and $\delta > 0$, RR-OPIM+ yields a $(\frac{1}{2} - \epsilon)$-approximate output $\mathcal{S} = \bigcup_{u \in A} \mathcal{S}_u$, with probability at least $1 - \delta$.*

Moreover, we have the following theorem to guarantee the expected running time of RR-OPIM+.

THEOREM 5.5. *With $\delta < 1/2$,* RR-OPIM+ *runs in the expected time of $O\left( \epsilon^{-2} \cdot \left( k \cdot d \cdot \ln|C| + \ln \frac{1}{\delta} \right) \cdot \left( n_p + m_p \cdot \frac{\sigma(\{v^*\})}{\sigma(\mathcal{S}^*)} \right) \right)$, where $\mathcal{S}^*$ is the optimal seed set, $v^*$ is the node with the largest spread in $P$.*

## 6 ADDITIONAL RELATED WORKS

In this part, we revisit problems that are germane to our work at first glance and distinguish them from in-game scenarios and CIM. Amid them, a line of works employs IM for link prediction. For example, active friending [53] and IM variants based on edge insertion [5, 10, 23] recommend people-you-may-know to increase the acceptance probability and boost the influence spread, respectively. In contrast, the in-game scenario is to sort existing friends for APs, and the objective of CIM can be fundamentally treated as the ranking measure. Lu et al. [31] focuses on the comparative IM problem and considers scenarios where two products are promoted simultaneously, which is beyond the scope of our study. Li et al. [27] leverages the Hawkes process to infer the diffusion, and it is orthogonal to our work. Prior works in [19, 42] consider adaptive seeding and assume that the seed selection is a two-stage framework, which first selects a set $S$ from a given subset of $V$, followed by selecting another seed set $T$ from the influenced neighboring nodes of $S$. The objective is to maximize the expected influence spread of $T$ under the cardinality constraint of $|S| + |T|$. Distinct from adaptive seeding, in-game incentive propagation only focuses on the second stage but requires a partition matroid constraint. The self-activation influence maximization [44] introduces a concept called self-activated user, which is similar to AP but also fails to consider the capacity of each AP. Another related work is multi-round influence maximization [45]. It considers the scenario requiring multiple rounds of promotions and may desire to repeatedly select the same influential user, whereas this contradicts Observation 2. Huang et al. [20] study influence maximization of online gaming platforms but also consider the summation of the influence of single seeds. In addition, targeted influence maximization [43] aims to select seeds to influence more users from a targeted subset, whereas CIM describes a reverse problem starting from a subset.

## 7 EXPERIMENTS

We first introduce the experimental settings, followed by conducting an empirical study on *TXG* to explore the configurations for CIM. At last, we evaluate the performance of the proposed algorithms in terms of quality and efficiency. All experiments are conducted on a Linux machine with Intel Xeon(R) Gold 6240@2.60GHz CPU and 377GB RAM in single-thread mode. None of the experiments need anywhere near all the memory. Due to space constraints, we refer interested readers to Appendix A for more experiments, e.g., AP selection and sensitivity analysis w.r.t. other constants.

Table 1: Dataset statistics ($K = 10^3$, $M = 10^6$, $B = 10^9$).

| Name | #nodes ($n$) | #edges ($m$) |
|---|---|---|
| DNC | 0.9$K$ | 24.2$K$ |
| Blog | 10.3$K$ | 668.0$K$ |
| Twitch | 168.1$K$ | 13.6$M$ |
| TXG | 243.4$K$ | 11.8$M$ |
| Orkut | 3.1$M$ | 234.2$M$ |
| Twitter | 41.7$M$ | 2.9$B$ |

## 7.1 Experimental Setups

**Datasets.** Besides the incentive propagation dataset *TXG*, we include 5 real-world social networks: *DNC* [24], *Blog* [48], *Twitch* [40], *Orkut* [54] and *Twitter* [55], whose statistics are shown in Table 1. All datasets are collected from KONECT [24] and SNAP [26], and used in previous IM works [17, 18, 49].

**Algorithms and parameters.** We test the performance of 9 algorithms, which are in the following 3 categories:
- *Local competitors:* Degree, PageRank, IMM, OPIM-C. We follow the local framework in Section 2.3 to select $k$ seeds for each AP by degree, PageRank [36], IMM [49], and OPIM-C [46].
- *Greedy solutions:* MG-Greedy, RR-Greedy. We utilize the lazy evaluation trick in CELF for the seed selection and conduct $r = 10,000$ MC simulations for spread estimation [25].
- *Scalable solutions:* RR-OPIM+, RR-OPIM, MG-OPIM. To verify the tightened bound in RR-OPIM+, we replace Eq.(4) with Eq.(2) in Line 6 of Algorithm 3 and call it RR-OPIM. To evaluate RR-Greedy in RR-OPIM+, we replace RR-Greedy with MG-Greedy in Line 5 of RR-OPIM and call it MG-OPIM. Notice that results of RR-OPIM and MG-OPIM are also $(1/2 - \epsilon)$-approximate.

For algorithms extending OPIM-C, we set $\epsilon = 0.1$ and $\delta = 1/n$ [17, 18, 46]. For a fair comparison, all algorithms are implemented in C++ and compiled with −03 optimization, which is available at: https://github.com/waetr/Capacity_Constrained_IM.

## 7.2 Empirical Configurations for CIM

**Selecting M.** First, we bridge the gap between model M (IC or LT) and the actual dissemination from the engaged seeds. To generate the diffusion among PPs, we use the co-playing logs with tuples $(u, v, T_{u,v})$ representing that an active PP $u$ played together with the passive friend $v$ at timestamp $T_{u,v}$, and clean logs by preserving the earliest timestamp for each distinct co-playing relationship. We construct the *diffusion trees* [38] from the co-playing logs. In particular, we treat the seed as the root of each tree and add the directed edge $(v, w)$ to the tree if (i) there exists an edge $(u, v)$ on the tree satisfying $T_{u,v} < T_{v,w}$ and (ii) the tree is acyclic after insertion. To capture the diffusion of co-playing behaviors, we follow prior works [22, 49, 50] to normalize the influence probability and weight in IC and LT by leveraging daily co-playing times between friends. Analogously, the model-predicted diffusion can also be preprocessed into a diffusion tree. Figure 2(a) reports the RMSE between the amount of predicted and true active PPs in each hop $t$, where the active users have the same shortest distance $t$ from the seed set $S$. We find that IC has a better RMSE on each hop, indicating that the actual diffusion is more similar to IC rather than LT and motivating us to leverage IC for CIM. To explain, the inactive $v$ acquires the incentive automatically once $v$ plays with one of its active friend $u$ for the first time, which resembles the process of IC.

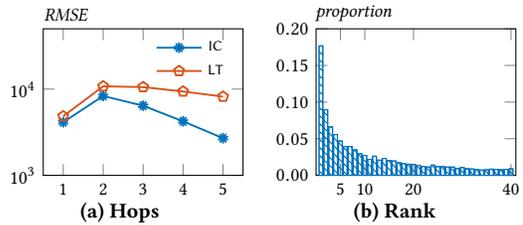

Figure 2: The RMSE of estimating spreads in each hop. The distribution of the rank of all invited passive seeds on *TXG*.

Table 2: The actual spread on *TXG*.

| Algorithm | RR-OPIM+ | MG-OPIM | RR-OPIM | Degree | PageRank |
|---|---|---|---|---|---|
| Spread | 1,632 | 1,625 | 1,609 | 1,488 | 1,471 |

**Selecting $k$.** We next explore the setting of $k$. Specifically, we collect each invited seed $v$'s rank (i.e., position) in the recommendation list of an AP $u$. Figure 2(b) reports the distribution of the ranks of all invited seeds. Notably, the AP prefers inviting the seeds in the top ranks. For instance, 80.1% of seeds are invited when they rank in the *top 20*. We also analyze the rank distribution w.r.t. the overlapped and invited seeds, which is similar to Figure 2(b), indicating the overlapping phenomenon is ubiquitous in the top positions. To summarize, an AP is more likely to invite seeds in the top 20 positions, where seeds are usually overlapping.

## 7.3 Performance Evaluation

In the second set of experiments, we compare the performance of each algorithm in terms of effectiveness and efficiency. Regarding the game dataset *TXG*, we retain the seeds that continued to daily contamination and their preceding APs, resulting in 794 APs and 1.7 thousand seeds. These 1.7 thousand seeds are treated as candidates, and each algorithm is asked to choose 1 seed for each of the 794 APs from their respective candidates. The actual spread of a subset $S$ of candidates is the number of PPs activated when those seeds are chosen. To configure CIM on remaining public datasets, we only leverage the fraction of APs as explored in Section 3.1, and uniformly sample 5% fraction of users from $V$ as the AP set $A$, i.e., $d/n = 5\%$. After determining $A$, we leverage IC as M on the derived subgraph $P$ and choose the constant $k$ ranging from 2 to 20. We treat $A$ and the corresponding $P$ as a query set, and report the average score after repeating on 5 random query sets. We exclude an algorithm if it fails to return in 24 hours.

**Effectiveness analysis.** Table 2 shows the actual spread of the $S$ selected by each method (the timeout one is omitted), where the proposed RR-OPIM+ outperforms other methods by up to 9.87%. Figure 3 reports the spread of each approach on public datasets by fixing $d/n = 5\%$ and varying $k$. Regarding MG-Greedy and RR-Greedy, RR-Greedy is slightly better than MG-Greedy, and both greedy solutions are superior to other solutions on *DNC*. Notably, the seeds of RR-Greedy infect 26.8% more PPs than Degree when $k = 10$. Regarding the scalable solutions, we find that their seed qualities are comparable to greedy solutions and better than local competitors. For instance, RR-OPIM+ can improve the local heuristics by up to 39% on *Orkut*. Regarding the local competitors, they are inferior to all greedy solutions and their scalable versions due to the overlapping of seeds mentioned in Section 2.2. In particular, we report the number of distinct seeds returned by different solutions in Figure 4, where the seed size of RR-OPIM+ is at least 3× larger than Degree. Furthermore, we observe that OPIM-C and IMM have better results than the rest local solutions on *DNC*, showing the usefulness of approximation in the local framework. We also report the spread by fixing $k = 10$ and varying $d/n \in \{5, 10, 20, 50\}‰$. As shown in Figure 5, the results under different $d$ settings have the same tendency as those under different $k$ settings.

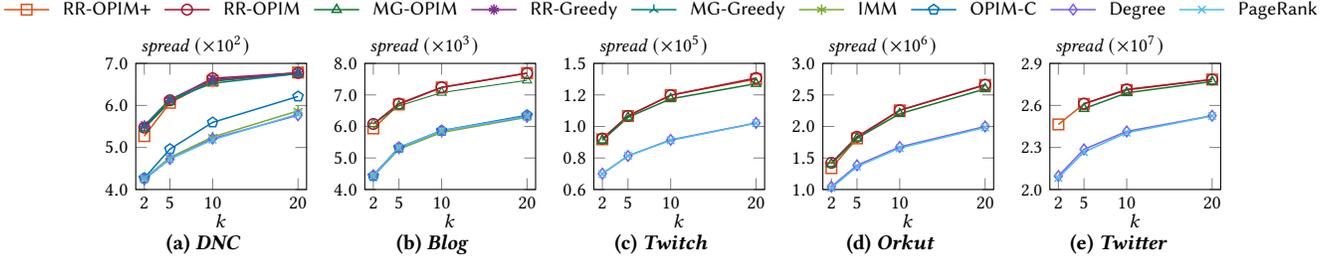

Figure 3: The spread on various graphs by varying $k$ ($d/n = 50‰$).

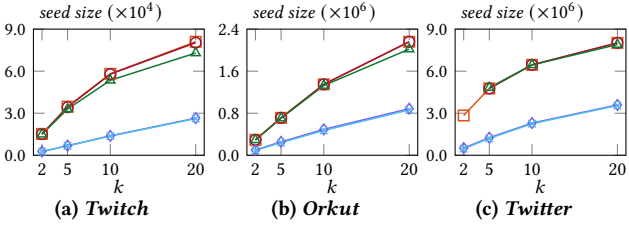

Figure 4: The seed size by varying $k$ ($d/n = 50‰$).

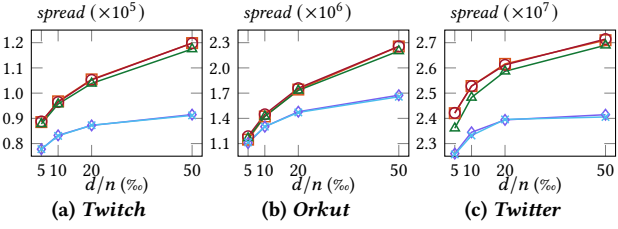

Figure 5: The spread by varying $d$ ($k = 10$).

**Efficiency analysis.** We next compare the running time of each solution. Here, we ignore the running time of Degree, which can be recorded while reading the graph. As illustrated in Figure 6 and Figure 7, the scalable solution RR-OPIM+ outperforms other solutions in all cases. Most notably, RR-OPIM+ improves RR-OPIM and MG-OPIM by two orders of magnitude on *Twitch*, which signifies the superiority of the proposed tightened bound. Furthermore, we find that RR-OPIM costs less time than MG-OPIM, which spends more time in seed selection and requires more iterations. For instance, MG-OPIM is about 2× slower than RR-OPIM on *Orkut* when $k \geq 10$. For greedy solutions, due to the inefficiency of MC simulations, both solutions are only feasible on *DNC* but fail on the rest. Notice that the running time of MG-Greedy is slightly better than RR-Greedy. This might be caused by the pruning procedure in CELF. We also evaluate the running time on a smaller co-authorship graph with 1.5K nodes and 2.7K edges [24], where RR-Greedy is up to 200× faster than MG-Greedy by varying $d$. Regarding other local competitors, they cost much more time than RR-OPIM+. For instance, OPIM-C is 3 to 4 orders of magnitude slower than RR-OPIM+ on *DNC* and *Blog*. PageRank is about one order of magnitude slower than RR-OPIM+ on *Twitch*, *Orkut*, and *Twitter*. Moreover, IMM and OPIM-C raise a timeout on *Twitch*, *Orkut*, and *Twitter*, due to $O(d)$ times of invocations.

**Sensitivity Analysis.** At last, we explore how the error constant $\epsilon$ affects scalable implementations. Figure 8 reports the time-spread curve by fixing $k = 10, d/n = 5\%$ and varying $\epsilon \in \{0.1, 0.05, 0.02, 0.01\}$. The results are sorted in the descending order of $\epsilon$, where RR-OPIM and MG-OPIM has a timeout when $\epsilon = 0.01$ in Figure 8(b) and they only have one point when $\epsilon = 0.1$ in Figure Figure 8(c). As shown, the running time and spread of RR-OPIM+ are fast and robust when $\epsilon$ varies. In contrast, the running time of RR-OPIM and MG-OPIM is more sensitive to $\epsilon$. In particular, as $\epsilon$ is halved, the running time of RR-OPIM and MG-OPIM increases by $2\times -10\times$.

## 8 DEPLOYMENTS

We have deployed RR-OPIM+ on an incentive propagation event of Tencent's battle royale game X with 88.2 million quarter-active users and 3.2 billion relationships, whose procedure follows the description in Section 2.3. This deployment is conducted on an in-house cluster consisting of hundreds of machines, each of which has 16GB memory and 12 Intel Xeon Processor E5-2670 CPU cores.

In the friendship network of X, the weight of each edge is described by the *intimacy* score, which records the number of historical interactions from one to the other, e.g., co-playing, gifting, and so on. We implement our proposal by first invoking RR-OPIM+ to select $k$ passive seeds for each AP and then ranking each seed in the descending order of its intimacy value with the AP. This ensures the interaction willingness between APs and the selected seeds. For the sake of fairness, we compare the performance of our proposal with the strategy Intimacy, which directly ranks friends by intimacy scores and is widely accepted by in-game friend ranking [57]. Each approach is initially computed based on the subgraph instance ahead of the event and is then updated daily by using the latest graph snapshot. The APs are selected according to the active week, returning 373.46 thousand and 382.52 thousand APs for our proposal and Intimacy, respectively. Due to the network effect, we follow [41] and partition all users into communities with high connectivity and feature similarity. We then conduct the online A/B testing that randomly assigns the live traffic in the same community to the treatment (i.e., ours) or control (i.e., Intimacy) group.

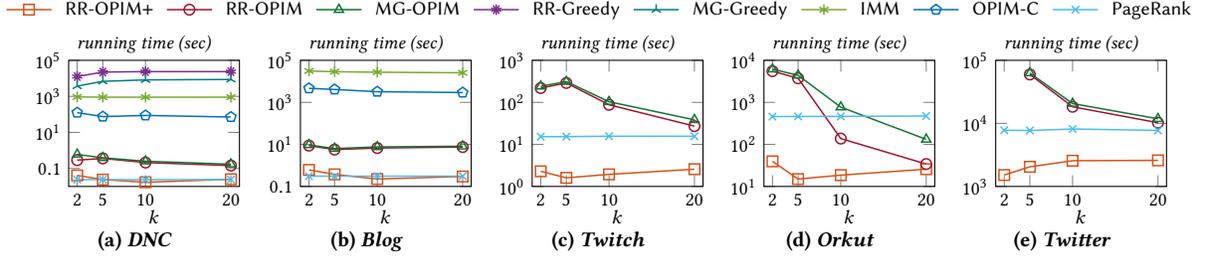

Figure 6: The running time on various graphs by varying $k$ ($d/n = 50‰$).

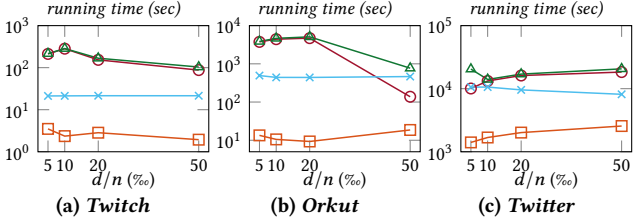

Figure 7: The running time by varying $d$ ($k = 10$).

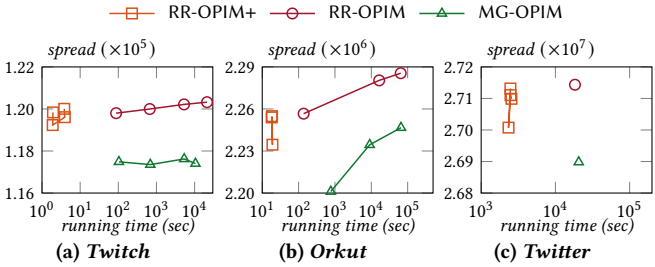

Figure 8: Running time vs. spread by varying $\epsilon$.

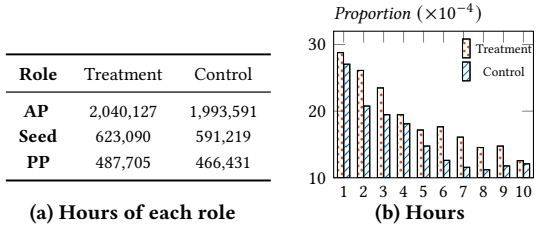

(a) Hours of each role　　(b) Hours

Figure 9: The playing hours of active users in each role and the distribution of playing hours of each active user on X.

When the event ended, we evaluate the performance based on the *total spread*, i.e., the number of APs and seeds engaged in seed activation, as well as the activated PPs in daily contamination, which is 60.69 thousand for the treatment group and 58.28 thousand for the control group. This improvement is statistically significant, and we refer interested readers to Appendix A. We also evaluate the playing hours of the total spread and break them down based on roles, i.e., AP, seed, and PP. As shown in Figure 9(a), we can find that the treatment group yields more playing hours for all user roles. Notably, the treatment group improves the control group by 2.33%, 5.39%, and 4.56% for APs, seeds, and PPs, respectively. In addition, we measure the distribution of each active user's playing hours. Figure 9(b) reports the distribution of playing hours ranging from 1 to 10. As shown in Figure 9(b), we can observe that (i) for both the treatment and control groups, the number of active users decreases as the playing hour increases, and (ii) our proposal attracts more users than the baseline in all playing hours. It is worth noting that a similar result can be found for the remaining hours.

## 9 CONCLUSIONS

Motivated by in-game insights, we present CIM and offer two greedy solutions MG-Greedy and RR-Greedy. For the sake of scalability, we further design an approximated algorithm RR-OPIM+ with near-linear running time. We conduct extensive experiments to demonstrate the superiority of our proposal in terms of effectiveness and efficiency. In addition, we deploy RR-OPIM+ to the in-game incentive propagation scenario, achieving considerable improvement. As a future direction, we will aim to improve the approximation ratio of our proposal and consider the dynamic setting.

## 10 ETHICS STATEMENT

While the proposed CIM problem and its solutions may boost user engagement and revenue for the online platform, they also pose privacy risks due to the collection of user data and may expose users to negative side effects, such as addiction. As researchers, we recognize the importance of balancing the potential benefits of studying these algorithms with the potential risks to user well-being. To ensure privacy and confidentiality, we have strictly anonymized the data used in our study and have no access to detailed user profiles. Additionally, we have followed the ethical guidelines [37] set forth by Tencent Inc. in conducting this research.

## A APPENDIX

### A.1 Proofs

**Proof of Corollary 1.** For any graph $G = (V, E)$, we first construct an extended graph $G'$ by adding a node $x$ to $V$ and directed edges $e_{x,v}$ for all $v \in V$ to $E$. Notice that the IM problem on $G$ is NP-hard [22] and is the special case of the CIM problem on $G'$ with $A = \{x\}$. Hence, the CIM problem is also NP-hard. □

**Proof of Theorem 4.3.** Denote $e_{u_i,v_i}$ be the $i$-th selected AP-seed edge by Algorithm 2, where $1 \leq i \leq b$ and $b = d \cdot k$ without loss of generality. Let $S_i$ be the set of first $i$ selected pairs $\{e_{u_1,v_1}, e_{u_2,v_2}, \ldots, e_{u_i,v_i}\}$ for $i \geq 1$, where $S_0 = \emptyset$ and $S_b = S$ is the output set of Algorithm 2. We let $S_i^*$ be the seed set with maximal spread which contains $S_i$ and satisfies the partition matroid constraint. By Algorithm 2, we have $S_0^* = S^*$ and $S_b^* = S_b = S$. For $i = 1, \ldots, b$, let $u$ be the AP considered at Line 4 and $e_{u,v}$ be the pair adding to $S_{i-1}$, i.e., $S_i = S_{i-1} \cup \{e_{u,v}\}$. Since $S_{i-1} \subset S_{i-1}^*$ and $|S_{i-1} \cap C_u| < k$, $(S_{i-1}^* \setminus S_{i-1}) \cap C_u \neq \emptyset$. Let $e_{u,v'}$ be an arbitrary pair in $(S_{i-1}^* \setminus S_{i-1}) \cap C_u$. Since $\sigma(S_i^*) \geq \sigma(S_{i-1}^* \cup \{e_{u,v}\} \setminus \{e_{u,v'}\})$,

we have

$$\sigma(\mathcal{S}_{i-1}^*) - \sigma(\mathcal{S}_i^*) \leq \sigma(\mathcal{S}_{i-1}^*) - \sigma(\mathcal{S}_{i-1}^* \cup \{e_{u,v}\} \setminus \{e_{u,v'}\})$$
$$= \left[\sigma(\mathcal{S}_{i-1}^*) - \sigma(\mathcal{S}_{i-1}^* \setminus \{e_{u,v'}\})\right] - \qquad (5)$$
$$\quad \left[\sigma(\mathcal{S}_{i-1}^* \cup \{e_{u,v}\} \setminus \{e_{u,v'}\}) - \sigma(\mathcal{S}_{i-1}^* \setminus \{e_{u,v'}\})\right]$$
$$= \sigma(e_{u,v'}|\mathcal{S}_{i-1}^* \setminus \{e_{u,v'}\}) - \sigma(e_{u,v}|\mathcal{S}_{i-1}^* \setminus \{e_{u,v'}\}).$$

Since $\mathcal{S}_{i-1} \subseteq \mathcal{S}_{i-1}^* \setminus \{e_{u,v'}\}$, by the submodularity of $\sigma(\cdot)$ and the definition of $\gamma$, Eq. (5) is bounded by

$$\sigma(\mathcal{S}_{i-1}^*) - \sigma(\mathcal{S}_i^*) \leq \sigma(e_{u,v'}|\mathcal{S}_{i-1}) - (1-\gamma)\sigma(e_{u,v}|\mathcal{S}_{i-1}).$$

Recall that $e_{u,v}$ is the element obtained in Algorithm 2, i.e.,

$$\sigma(e_{u,v}|\mathcal{S}_{i-1}) \geq \sigma(e_{u,v'}|\mathcal{S}_{i-1})$$

for all $e_{u,v'} \in C_u \setminus \mathcal{S}_{i-1}$. Therefore,

$$\sigma(\mathcal{S}_{i-1}^*) - \sigma(\mathcal{S}_i^*) \leq \sigma(e_{u,v}|\mathcal{S}_{i-1}) - (1-\gamma) \cdot (e_{u,v}|\mathcal{S}_{i-1})$$
$$= \gamma \cdot (\sigma(\mathcal{S}_i) - \sigma(\mathcal{S}_{i-1})).$$

Hence, we can obtain that

$$\sigma(\mathcal{S}^*) - \sigma(\mathcal{S}_b^*) = \sum_{t=1}^{b}(\sigma(\mathcal{S}_{t-1}^*) - \sigma(\mathcal{S}_t^*)) \qquad (6)$$
$$\leq \gamma \cdot \sum_{t=1}^{b}(\sigma(\mathcal{S}_t) - \sigma(\mathcal{S}_{t-1})) \leq \gamma \cdot \sigma(\mathcal{S}_b).$$

Since $\mathcal{S}_b^* = \mathcal{S}_b = \mathcal{S}$, Eq. (6) turns to

$$\sigma(\mathcal{S}^*) \leq (1+\gamma) \cdot \sigma(\mathcal{S}),$$

which completes the proof. □

In what follows, we mainly utilize two martingale-based concentration bounds from [49], which intuitively describes how close between $\Lambda_{\mathcal{R}}(\cdot)$ and $\frac{\theta}{n} \cdot \sigma(\cdot)$ for a given number of random RR sets.

**Lemma A.1 ([49]).** Given a graph $G$ with $n$ nodes, a set $\mathcal{R}$ of random RR sets with $|\mathcal{R}| = \theta$ and a seed set $S$, for any $\lambda > 0$,

$$\Pr\left[\Lambda_{\mathcal{R}}(S) - \sigma(S) \cdot \tfrac{\theta}{n} \geq \lambda\right] \leq \exp\left(-\frac{\lambda^2}{2 \cdot \sigma(S) \cdot \frac{\theta}{n} + \frac{2}{3}\lambda}\right), \quad (7)$$

$$\Pr\left[\Lambda_{\mathcal{R}}(S) - \sigma(S) \cdot \tfrac{\theta}{n} \leq -\lambda\right] \leq \exp\left(-\frac{\lambda^2}{2 \cdot \sigma(S) \cdot \frac{\theta}{n}}\right). \quad (8)$$

**Proof of Lemma 5.1.** Recall that $n_p \cdot \frac{\Lambda_{\mathcal{R}}(\mathcal{S})}{\theta}$ is an unbiased estimator for $\sigma(\mathcal{S})$, where $|\mathcal{R}| = \theta$. Hence, a large enough $\theta$ can make this estimator close to $\sigma(\mathcal{S})$. This connection is shown by the following lemma.

**Lemma A.2 ([49]).** Given a graph $P$ with $n_p$ nodes, constants $\epsilon_1$ and $\delta_1$, if $\theta \geq \theta_1 = \frac{2 \cdot n_p \cdot \ln \frac{1}{\delta_1}}{\sigma(\mathcal{S}^*) \cdot \epsilon_1^2}$, then

$$\Pr\left[n_p \cdot \tfrac{\Lambda_{\mathcal{R}}(\mathcal{S}^*)}{\theta} \geq (1-\epsilon_1) \cdot \sigma(\mathcal{S}^*)\right] \geq 1 - \delta_1.$$

Combining Corollary 2 and Lemma A.2, we can derive that following Eq. (9) hold with probability at least $1 - \delta_1$ if $\theta \geq \theta_1$:

$$n_p \cdot \tfrac{\Lambda_{\mathcal{R}}(\mathcal{S})}{\theta} \geq \tfrac{n_p}{2} \cdot \tfrac{\Lambda_{\mathcal{R}}(\mathcal{S}^*)}{\theta} \geq \tfrac{(1-\epsilon_1)}{2} \cdot \sigma(\mathcal{S}^*). \quad (9)$$

We now need the following lemma to connect $\sigma(\mathcal{S})$ with $\sigma(\mathcal{S}^*)$.

**Lemma A.3.** Given a graph $P$ with $n_p$ nodes, constants $\epsilon_2 = \epsilon - \frac{\epsilon_1}{2} \geq 0$ and $\delta_2$, if Eq. (9) holds and

$$\theta \geq \theta_2 = \frac{n_p \cdot \left(\ln\left(\prod_{u \in A}\binom{|C_u|}{k}\right) + \ln\frac{1}{\delta_2}\right)}{\sigma(\mathcal{S}^*) \cdot \epsilon_2^2},$$

then

$$\Pr\left[\sigma(\mathcal{S}) \geq (\tfrac{1}{2} - \epsilon) \cdot \sigma(\mathcal{S}^*)\right] \geq 1 - \delta_2.$$

According to Lemmas A.2-A.3 and the union bound, if $\theta \geq \max(\theta_1, \theta_2)$,

$$\sigma(\mathcal{S}) \geq (\tfrac{1}{2} - \epsilon) \cdot \sigma(\mathcal{S}^*)$$

holds with the probability at least $1 - \delta_1 - \delta_2$. By setting

$$\epsilon_1 = \frac{\epsilon}{\left(\tfrac{1}{2} + \sqrt{\frac{\ln\left(\prod_{u \in A}\binom{|C_u|}{k}\right) + \ln\frac{1}{\delta_2}}{2\ln\frac{1}{\delta_1}}}\right)}$$

and $\delta_1 = \delta_2 = \delta/6$, we have

$$\theta_1 = \theta_2 = \frac{2n_p \cdot \left(\tfrac{1}{2}\sqrt{\ln\frac{6}{\delta}} + \sqrt{\tfrac{1}{2} \cdot \left(\ln\left(\prod_{u \in A}\binom{|C_u|}{k}\right) + \ln\frac{6}{\delta}\right)}\right)^2}{\epsilon^2 \cdot \sigma(\mathcal{S}^*)}.$$

Since $\chi \leq \sigma(\mathcal{S}^*)$ and $\theta_{max} \geq \max(\theta_1, \theta_2)$ by definition, this lemma holds with probability at least $1 - \delta/3$ if $\theta > \theta_{max}$, which completes the proof. □

**Proof of Lemma A.3.** Given a graph $P$ with $n_p$ nodes and any $\mathcal{S}' \in \mathcal{I}$, we say $\mathcal{S}'$ is bad if $\sigma(\mathcal{S}') < (\tfrac{1}{2} - \epsilon) \cdot \sigma(\mathcal{S}^*)$. Based on Eq. (7), for a bad $\mathcal{S}'$, we have

$$\Pr\left[n_p \cdot \tfrac{\Lambda_{\mathcal{R}}(\mathcal{S}')}{\theta} - \sigma(\mathcal{S}') \geq \epsilon_2 \cdot \sigma(\mathcal{S}^*)\right]$$
$$\leq \exp\left(-\frac{\epsilon_2^2 \cdot \sigma(\mathcal{S}^*)^2}{2 \cdot \sigma(\mathcal{S}') + \tfrac{2}{3} \cdot \epsilon_2 \cdot \sigma(\mathcal{S}^*)} \cdot \frac{\theta}{n_p}\right)$$
$$\leq \exp\left(-\frac{\epsilon_2^2 \cdot \sigma(\mathcal{S}^*)}{2 \cdot (\tfrac{1}{2} - \epsilon) + \tfrac{2}{3} \cdot \epsilon_2} \cdot \frac{\theta}{n_p}\right)$$
$$\leq \exp\left(-\frac{\epsilon_2^2 \cdot \sigma(\mathcal{S}^*) \theta_2}{n_p}\right)$$
$$\leq \delta_2 \bigg/ \left(\prod_{u \in A}\binom{|C_u|}{k}\right).$$

Since there exists at most $\prod_{u \in A}\binom{|C_u|}{k}$ possible $\mathcal{S}'$, by the union bound and Eq. (9), we obtain that, for any $\mathcal{S}'$ returned by Algorithm 3,

$$\sigma(\mathcal{S}') \leq n_p \cdot \tfrac{\Lambda_{\mathcal{R}}(\mathcal{S}')}{\theta} - \epsilon_2 \cdot \sigma(\mathcal{S}^*) \leq (\tfrac{1}{2} - \epsilon) \cdot \sigma(\mathcal{S}^*)$$

with the probability at most $\delta_2$, which completes the proof. □

**Proof of Lemma 5.2.** Regarding the upper bound $\sigma^u(\mathcal{S}^*)$, in terms of Corollary 2 and Eq. (8) in Lemma A.1, we have

$$\Pr\left[\sigma(\mathcal{S}^*) > \left(\sqrt{2 \cdot \Lambda_{\mathcal{R}_1}(\mathcal{S}) + \frac{\ln \frac{1}{p_f}}{2}} + \sqrt{\frac{\ln \frac{1}{p_f}}{2}}\right)^2 \cdot \frac{n_p}{\theta_1}\right]$$

$$\leq \Pr\left[\sigma(\mathcal{S}^*) > \left(\sqrt{\Lambda_{\mathcal{R}_1}(\mathcal{S}^*) + \frac{\ln \frac{1}{p_f}}{2}} + \sqrt{\frac{\ln \frac{1}{p_f}}{2}}\right)^2 \cdot \frac{n_p}{\theta_1}\right]$$

$$\leq \Pr\left[\left(\sqrt{\frac{\theta_1 \cdot \sigma(\mathcal{S}^*)}{n_p}} - \sqrt{\frac{\ln \frac{1}{p_f}}{2}}\right)^2 > \Lambda_{\mathcal{R}_1}(\mathcal{S}^*) + \frac{\ln \frac{1}{p_f}}{2}\right]$$

$$= \Pr\left[-\sqrt{2 \cdot \ln \frac{1}{p_f} \cdot \frac{\theta_1 \cdot \sigma(\mathcal{S}^*)}{n_p}} > \Lambda_{\mathcal{R}_1}(\mathcal{S}^*) - \frac{\theta_1 \cdot \sigma(\mathcal{S}^*)}{n_p}\right]$$

$$\leq \exp\left(-\frac{2 \cdot \ln \frac{1}{p_f} \cdot \frac{\theta_1 \cdot \sigma(\mathcal{S}^*)}{n_p}}{2 \cdot \frac{\theta_1 \cdot \sigma(\mathcal{S}^*)}{n_p}}\right)$$

$$= p_f.$$

The proof of the lower bound $\sigma^l(\mathcal{S})$ can be found in Lemma 4.2 of [46]. We show the proof here to make it self-contained. Specifically, by Eq. (7) in Lemma A.1, we have

$$\Pr\left[\sigma(\mathcal{S}) < \left(\left(\sqrt{\Lambda_{\mathcal{R}_2}(\mathcal{S}) + \frac{2 \cdot \ln \frac{1}{p_f}}{9}} - \sqrt{\frac{\ln \frac{1}{p_f}}{2}}\right)^2 - \frac{\ln \frac{1}{p_f}}{18}\right) \cdot \frac{n_p}{\theta_2}\right]$$

$$= \Pr\left[\sqrt{\frac{\sigma(\mathcal{S}) \cdot \theta_2}{n_p} + \frac{\ln \frac{1}{p_f}}{18}} < \sqrt{\Lambda_{\mathcal{R}_2}(\mathcal{S}) + \frac{2 \cdot \ln \frac{1}{p_f}}{9}} - \sqrt{\frac{\ln \frac{1}{p_f}}{2}}\right]$$

$$= \Pr\left[\sqrt{2 \cdot \ln \frac{1}{p_f} \cdot \frac{\sigma(\mathcal{S}) \cdot \theta_2}{n_p} + \frac{\ln^2 \frac{1}{p_f}}{9}} + \frac{\ln \frac{1}{p_f}}{3} < \Lambda_{\mathcal{R}_2}(\mathcal{S}) - \frac{\sigma(\mathcal{S}) \cdot \theta_2}{n_p}\right]$$

$$\leq \exp\left(-\frac{\left(\sqrt{2 \cdot \ln \frac{1}{p_f} \cdot \frac{\sigma(\mathcal{S}) \cdot \theta_2}{n_p} + \frac{\ln^2 \frac{1}{p_f}}{9}} + \frac{\ln \frac{1}{p_f}}{3}\right)^2}{2 \cdot \frac{\sigma(\mathcal{S}) \cdot \theta_2}{n_p} + \frac{2}{3} \cdot \left(\sqrt{2 \cdot \ln \frac{1}{p_f} \cdot \frac{\sigma(\mathcal{S}) \cdot \theta_2}{n_p} + \frac{\ln^2 \frac{1}{p_f}}{9}} + \frac{\ln \frac{1}{p_f}}{3}\right)}\right)$$

$$= p_f.$$

□

**Proof of Lemma 5.3.**

$$\Lambda_{\mathcal{R}}(\mathcal{S}^*) \leq \Lambda_{\mathcal{R}}(\mathcal{S}^* \cup \mathcal{S})$$

$$\leq \Lambda_{\mathcal{R}}(\mathcal{S}) + \sum_{e_{u,v} \in \mathcal{S}^* \setminus \mathcal{S}} \Lambda_{\mathcal{R}}(e_{u,v}|\mathcal{S})$$

$$= \Lambda_{\mathcal{R}}(\mathcal{S}) + \sum_{u \in A} \sum_{e_{u,v} \in \mathcal{S}^*_u \setminus \mathcal{S}_u} \Lambda_{\mathcal{R}}(e_{u,v}|\mathcal{S})$$

where the first two inequalities are derived from the monotonicity and submodularity of $\Lambda_{\mathcal{R}}(\cdot)$, respectively. Since for each $u \in A$, the number of $e_{u,v} \in \mathcal{S}^*_u \setminus \mathcal{S}_u$ does not exceed $k$, the above inequality can be further derived as

$$\Lambda_{\mathcal{R}}(\mathcal{S}^*) \leq \Lambda_{\mathcal{R}}(\mathcal{S}) + \sum_{u \in A} \sum_{e_{u,v} \in \Phi_k(\mathcal{S},u)} \Lambda_{\mathcal{R}}(e_{u,v}|\mathcal{S}),$$

which completes the proof. □

**Proof of Theorem 5.5.** Recall that RR-OPIM+ consists of (i) RR set generations, (ii) the node selection by RR-Greedy with the computation of $\sigma^u(\mathcal{S}^*)$ and $\sigma^l(\mathcal{S})$ in each iteration. In what follows, we analyze the time complexity of each subroutine in $i_{max}$ iterations.

Regarding the RR set generations, we follow Lemma 6.2 in [46] and provide the following corollary about the expected number of generated RR sets.

COROLLARY 3. *With $\delta \leq 1/2$, RR-OPIM+ generates an expected number of $O\left(\left(k \cdot d \cdot \ln |C| + \ln \frac{1}{\delta}\right) n_p \cdot \epsilon^{-2}/\sigma(\mathcal{S}^*)\right)$ RR sets.*

Combining Corollary 3 with Wald's equation [52] and the fact that the expected time of generating one RR set is bounded by $\frac{m_p}{n_p} \cdot \sigma(\{v^*\})$, where $v^*$ is the node with the largest spread in $P$ [50], we can derive that the time complexity for RR set generation is

$$O\left(\left(k \cdot d \cdot \ln |C| + \ln \frac{1}{\delta}\right) \cdot m_p \cdot \epsilon^{-2} \cdot \frac{\sigma(\{v^*\})}{\sigma(\mathcal{S}^*)}\right) \tag{10}$$

Regarding the rest subroutines in each iteration, RR-Greedy costs $O\left(\sum_{R \in \mathcal{R}_1} |R \cap C|\right)$ to scan all nodes $v$ in $C$. Moreover, RR-Greedy requires an additional time of $O(k \cdot |C|)$ for the computation of $\sigma^u(\mathcal{S}^*)$. Specifically, for each $0 \leq t < k$, $\Lambda^u_{\mathcal{R}_1}(\mathcal{S}^t)$ requires $O(|C_u|)$ time to choose the nodes in $\Phi_k(\mathcal{S}, u)$ for each $u \in A$. Akin to the seed selection in RR-Greedy, the overhead for computing $\sigma^l(\mathcal{S})$ is $O\left(\sum_{R \in \mathcal{R}_2} |R \cap C|\right)$, which further turns to

$$O\left(k \cdot |C| + \mathbb{E}[|\mathcal{R}_1 \cup \mathcal{R}_2|] \cdot \mathbb{E}[|R \cap C|]\right)$$

based on Wald's equation [52]. Note that the number of iterations $i_{max}$ is at most

$$i_{max} \leq \log_2\left(\frac{n_p}{\epsilon^2}\right) + 1 \leq \log_2(n_p) + \frac{2}{\epsilon} + 1.$$

Furthermore, let $\mathcal{R}_1$ and $\mathcal{R}_2$ are the sets of RR sets generated in the last iteration, then the total number of RR sets generated is no more than twice of $|\mathcal{R}_1 \cup \mathcal{R}_2|$. Therefore, the time complexity of rest subroutines in all iterations is

$$O\left(\left(\ln(n_p) + \epsilon^{-1}\right) \cdot k \cdot |C| + \mathbb{E}[|\mathcal{R}_1 \cup \mathcal{R}_2|] \cdot \mathbb{E}[|R \cap C|]\right).$$

LEMMA A.4. $\mathbb{E}[|R \cap C|] = \mathbb{E}[\sigma(\{u_c\})] \leq \sigma(\mathcal{S}^*)$, *where $u_c$ is a node selected uniformly at random from $C$.*

Based on Lemma A.4 and the above-said Corollary 3, the overall complexity of rest subroutines is

$$O\left(\left(\ln(n_p) + \epsilon^{-1}\right) \cdot k \cdot |C| + \left(k \cdot d \cdot \ln |C| + \ln \frac{1}{\delta}\right) n_p \cdot \epsilon^{-2}\right)$$

$$= O\left(\left(k \cdot d \cdot \ln |C| + \ln \frac{1}{\delta}\right) n_p \cdot \epsilon^{-2}\right). \tag{11}$$

Combining Eq.(10) and Eq.(11) derives the overall complexity of RR-OPIM+, which completes the proof. □

**Proof of Lemma A.4.** We denote the probability that a node set $S$ activates $v$ during the diffusion process as $Pr[S \to v]$. Notice that

$$\sigma(S) = \sum_{v \in V_p} Pr[S \to v].$$

Table 3: $\gamma_{max}$ and approximate ratio with different $d$ on *DNC*.

| $d$ | 5 | 10 | 20 | 50 |
|---|---|---|---|---|
| $\gamma_{max}$ | 0.931 | 0.970 | 0.978 | 0.983 |
| approx. rate | 0.518 | 0.508 | 0.506 | 0.504 |

Table 4: The fraction of engaged APs in the same active week.

| Active week | 1 | 2 | 3 | 4 | 5 | 6 |
|---|---|---|---|---|---|---|
| Engaged APs | 2.04% | 4.35% | 3.57% | 3.31% | 4.59% | 8.66% |

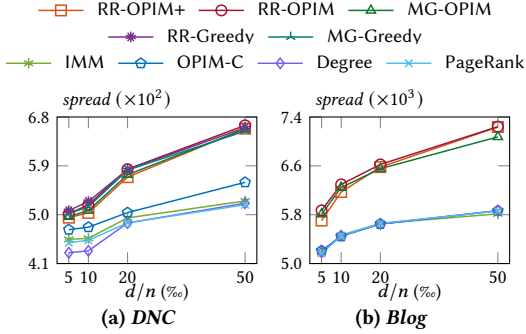

Figure 10: The spread by varying $d$ ($k = 10$).

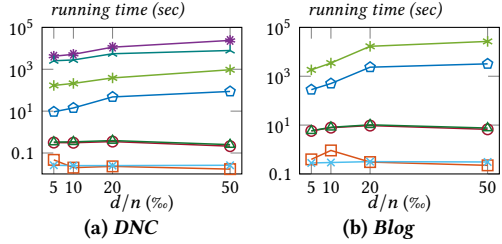

Figure 11: The running time by varying $d$ ($k = 10$).

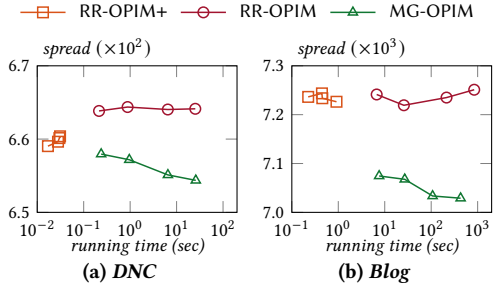

Figure 12: Running time vs. spread by varying $\epsilon$.

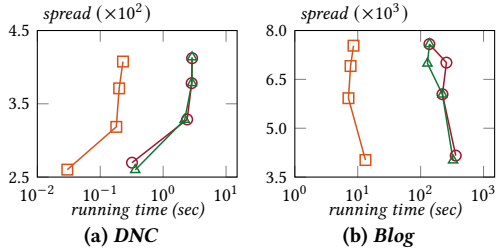

Figure 13: Running time vs. spread by varying $p_{u,v}$.

Therefore, the expected value of $|R \cap C|$ is equal to

$$\mathbb{E}[|R \cap C|] = \frac{1}{n_p} \sum_{v \in V_p} \sum_{u \in C} Pr[\{u\} \to v]$$
$$= \frac{1}{n_p} \sum_{u \in C} \sum_{v \in V_p} Pr[\{u\} \to v]$$
$$= \frac{1}{n_p} \sum_{u \in C} \sigma(\{u\})$$
$$= \mathbb{E}_{u \in C}[\sigma(\{u\})],$$

and according to the definition of $\sigma(S^*)$, it is trivial to get that $\mathbb{E}_{u \in C}[\sigma(\{u\})] \le \sigma(S^*) = \sigma(S^*)$. □

### A.2 More Results in Experiments and Deployments

**The values of $\gamma_{max}$.** Table 3 shows the values of $\gamma_{max}$ and the approximation ratio as $d$ increases on *DNC* as mentioned in Section 4.2. We conduct $r = 20,000$ MC simulations for spread estimation, with relative errors small enough to be considered as ground-truth. As $d$ increases, the candidate space $C$ is enlarged and the approximation ratio of RR-Greedy is close to 1/2.

**Selecting $A$.** We focus on understanding the correlation between the user's feature and the engagement of APs, which might be helpful for generating APs. Specifically, we employ a general feature called *active week*, and the active week $x$ of a user $u$ represents that $u$ keeps playing the game in the latest $x$ weeks. Table 4 reports the fraction of engaged APs over APs with the same active week, in which APs with the larger active weeks are more likely to join this event. Specifically, the Pearson correlation coefficient between the active week and AP engagement is 0.792, indicating these two features are strongly positive-correlated. Therefore, $A$ can be synthesized if users' historical activeness exists.

**Performance evaluation on *DNC* and *Blog*.** Figures 10-12 show the results on *DNC* and *Blog* in terms of result quality, running time, and the error constant $\epsilon$. The results share the same tendency as those on *Twitch*, *Orkut*, and *Twitter*, illustrating that RR-OPIM+ outperforms other solutions as $d$ varies. In addition, we conduct the sensitivity analysis w.r.t. the influence probability in the IC model. Specifically, we evaluate the performance of our proposed RR-OPIM+ and its two variants (RR-OPIM and MG-OPIM) by setting a uniform probability of $p_{u,v}$ on each edge and varying $p_{u,v} \in \{0.01, 0.04, 0.07, 0.1\}$ while keeping other parameters at default values ($\epsilon = 0.1$, $k = 10$, $d = 5\%$). Figure 13 reports the time-spread curve on DNC and Blog, showing that spread increases as $p_{u,v}$ increases. The reason is that increasing $p_{u,v}$ allows more nodes to be activated, resulting in more spread. In terms of running time, as the value of $p_{u,v}$ increases, more nodes are included in the RR set. However, fewer RR sets are needed to achieve the same approximation ratio. Consequently, constructing a single RR set may take more time, but the overall running time could be reduced.

**Significance analysis.** As explained, prior to the event, we first partitioned the social network into different clusters to mitigate network interference, and then randomly assigned clusters of users to the control and treatment groups, resulting in 11,927,377 control group users ($N_c$) and 11,465,639 treatment group users ($N_t$). Based on each group's total population, we selected the top 3% of users as APs according to their historical activeness and conducted the A/B test. Upon the event's conclusion, 58,280 users in the control group ($E_c$) and 60,691 users in the treatment group ($E_t$), comprising APs, seeds, and other activated PPs, participated and earned shared incentives. Using these statistics, we calculated the engagement probability for users in the control group ($p_c$) as $E_c/N_c = 0.489\%$ and in the treatment group ($p_t$) as $E_t/N_t = 0.529\%$. This demonstrates an 8.33% improvement in the treatment group compared to the control group. To assess the statistical significance of this improvement, we conducted the Z-test, a widely applied hypothesis test for discerning significant differences between two population means, especially in large populations, and A/B testing contexts for recommendation strategies' outcomes [56]. As a result, we obtained a Z-score of 13.825 using this formula:

$$z = \frac{p_t - p_c}{\sqrt{\frac{p_t \cdot (1-p_t)}{N_t} + \frac{p_c \cdot (1-p_c)}{N_c}}} = 13.825.$$

The corresponding two-tailed p-value is less than 0.01, signifying that the improvement is statistically significant.